# Transport Properties of AB stacked (Bernal) Bilayer Graphene on and without Substrate within 2- and 4-band Approximations

V.P. Gusynin[1,b)], S.G. Sharapov[1,c)] and A.A. Reshetnyak[2,a)]

[1]*Bogolyubov Institute for Theoretical Physics, 14-b, Metrolohichna str., Kiev, 03680 Ukraine*
[2]*Institute of Strength Physics and Materials Science SB RAS, 2/4, pr. Akademicheskii, Tomsk, 634021, Russia.*

a) Corresponding author: reshet@ispms.tsc.ru
b) vgusynin@bitp.kiev.ua
c) sharapov@bitp.kiev.ua

**Abstract.** We present the results of the calculations of longitudinal and Hall conductivities of AB-stacked bilayer graphene as a function of frequency, finite chemical potential, temperature both with and without magnetic fields on a base of 2- and 4-band effective models. The limited cases of the conductivities for direct current are derived. The relations being important for optoelectronic among Hall conductivities and Faraday, Kerr angles in the AB-bilayers samples in the electric and magnetic fields when the radiation passes across bilayer sheets on different kinds a substrate are obtained.

## INTRODUCTION

Graphene bilayer since its discovery [1] presents a unique representative of 2D condensed matter systems possessing by outstanding mechanical and transport properties. Its low-energy electron spectrum [2,3] combines characteristics of monolayer graphene and usual 2D electron systems. It consists of two inequivalent pairs of parabolic valence and conductance bands, touching each other at $K_+(K)$ and $K_-(K')$ Dirac points with massive and chiral charge carriers. One of the basic peculiarities of AB-bilayer concerns in the fact that an electric field $E$ applied perpendicular to the sheets leads to the opening of a tunable gap between the valence and conduction bands (see, Figure 1, where $b_1 = 2\pi/a(1,1/\sqrt{3})$, $b_2 = 2\pi/a(1,-1/\sqrt{3})$, $a=2,46$Å; $a_{cc}=a/\sqrt{3}=1,42$Å appear respectively by the primitive reciprocal lattice vectors, distances among adjacent unit cells and atoms, whereas in the vicinity of the $K_+(K)$ point the dashed square on Figure 1(b) marks out the 4-band energy spectrum).

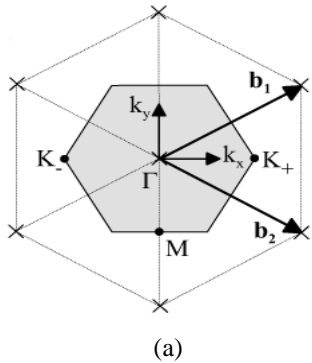
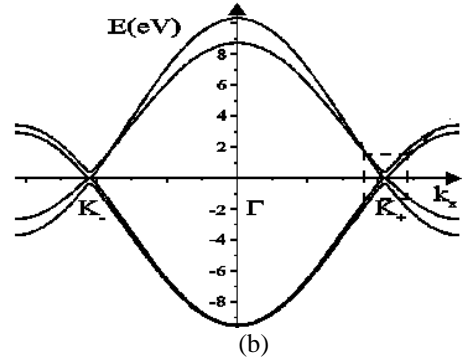

(a)   (b)

**FIGURE 1.** Reciprocal lattice of bilayer graphene in (a) with lattice points indicated as crosses, the shaded hexagon being by the first Brillouin zone with the Dirac points $K_-(K')$, $K_+(K)$ − showing two non-equivalent corners. (b) Low-energy bands of bilayer graphene arising from $2p_z$ orbitals plotted along the $k_x$ axis in reciprocal space intersecting the Dirac points. Plots were adapted from [3].

Note, the value of $E$ can be controlled externally by chemical doping and gating. Since, in contrast to single graphene, the density of states remains finite even in the unbiased and neutral bilayer, there are theoretical predictions [5] that the electron-electron interaction can result in spontaneous symmetry breaking and opening a gap even without a magnetic field. The nature of the gapped state is much debated in the literature. Possible scenarios include anomalous quantum Hall (QAH), quantum spin Hall (QSH), layer antiferromagnet (LAF) states, etc. Formally, all these gapped states are differed by the variant of breaking an approximate $SU(4)$ spin-valley symmetry of the low energy Hamiltonian of bilayer graphene.

The problem of constructing a field theoretical model describing an energy spectrum and optical- and magneto-optical conductivities of AB- bilayer in external electromagnetic fields for non-zero temperature and densities of charged carriers is not yet completely solved, (see, the results on optical conductivity in 4-band model [6]).

The paper is organized as follows. In the next Section we show how to obtain from tight-binding approach the Hamiltonian in the continuum approximation. Then we calculate the longitudinal and transverse (Hall) optical conductivities and study the choice of the different gapped states on it among them in magnetic field as well. The relations among the Hall conductivities and Faraday and Kerr angles when bilayer may be interacting with substrates are analyzed in the fourth Section. In the final Section we make concluding remarks.

## MODEL HAMILTONIANS FOR AB-STACKED BILAYER

The tight-binding model developed for graphite [7,8] and known as Slonczewski–Weiss–McClure (SWC) model are easily formulated for Hamiltonian of AB-stacked graphene according to hopping parameters in the Figure 2 (b) as

$$H_{tb} = -\gamma_0 \sum_{(i,j),m,s=1,2} a^+_{m,i,s} b_{m,j,s} - \gamma_4 \sum_{i;s=1,2}\left(a^+_{1,i,s} a_{2,i,s} + b^+_{1,i,s} b_{2,i,s}\right) - \gamma_1 \sum_{i;s=1,2} a^+_{2,i,s} b_{1,i,s} - \gamma_3 \sum_{i;s=1,2} a^+_{1,i,s} b_{2,i,s} + h.c., \quad (1)$$

where $a^+_{m,i,s}$ $(b^+_{m,j,s})$ creates an electron with spin s = $-\hbar/2$, $\hbar/2$ in layer m=1,2 on sublattice A(B) at site $\mathbf{R}_i$ with hopping parameters $(\gamma_0, \gamma_1, \gamma_3, \gamma_4)$ = (3.0, 0.38, 0.1, 0.12) eV with distance $d_0$=3.35Å between the layers [4]

Note, the atoms $A1$ and $B1$ on the lower layer in the Figure 2 are shown as white and black dots, $A2$, $B2$ on the upper layer are black and grey, correspondingly. The shaded rhombus in (*a*) indicates the conventional unit cell with .4 atoms A1, B1 (on the lower layer), A2, B2.(on the upper layer). The layers are arranged so that one of the atoms from the lower layer $B1$ is directly below an atom, $A2$, from the upper layer known as "dimer" sites because the electronic orbitals on them are coupled together by a relatively strong interlayer coupling.

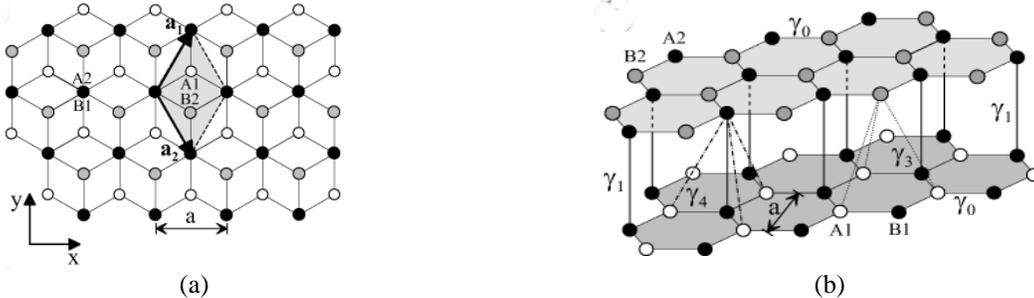

(a) (b)

**FIGURE2** (*a*) view in a perpendicular direction to the bilayer sheets and (*b*) side view of the crystal structure of bilayer graphene.

In the continuum limit, to study the electronic properties, when expending the momentum $p = \hbar(k + K_\xi)$ in the vicinity of K, K' points of fixed unit sell of the bilayer we obtain the effective 4-band Hamiltonian with 4-component spinor wave function $\Psi$, with general form for the gap, $\Delta_{\xi s}$, containing valley and spin depending terms [9]

$$H = \xi \begin{pmatrix} \Delta_{\xi s} & \gamma_3 v_F \pi/\gamma_0 & \gamma_4 v_F \pi^+/\gamma_0 & v_F \pi^+ \\ \gamma_3 v_F \pi^+/\gamma_0 & -\Delta_{\xi s} & v_F \pi & \gamma_4 v_F \pi/\gamma_0 \\ \gamma_4 v_F \pi/\gamma_0 & v_F \pi^+ & -\Delta_{\xi s} & \gamma_1 \\ v_F \pi & \gamma_4 v_F \pi^+/\gamma_0 & \gamma_1 & \Delta_{\xi s} \end{pmatrix} \text{ for } \Delta_{\xi s} = U + \xi \Delta_T + s U_T + \xi s \Delta \quad (2)$$

with Fermi velocity $v_F = 10^6\,m/s$ in plane of layers, $U$, $\Delta$ ($U_T$, $\Delta_T$) being by (not) invariant with respect to the time reversal, whereas the first gap $U$ being, besides, by dynamically generated can be induced by a perpendicular electric field $E$. In this case $U=eEd_0/2$ with $-e$ being by the charge of electron. The quantity $\pi = p_x + ip_y$ in (2) may be enlarged, by means of preserving of the $U(1)$ gauge invariance for the Hamiltonian $H$ in case of bilayer interaction with external magnetic field $\mathbf{B}$ being applied perpendicular to the sheets along the positive axis z realizing the case of QED in (2+1)-dimensional space-time, to the quantity: $\hat{\pi} = \hat{p}_x + i\hat{p}_y = -i\hbar(D_x + iD_y)$ for covariant derivative $D_i = \partial_i + (ie/\hbar c)A_i$. We choose the vector potential $\vec{A}(x,y) = (0, Bx)$ in the Landau gauge. The Hamiltonian $H$ acts on a wave function $\Psi_4$ corresponding to the atomic sites $A1$, $B2$, $A2$, $B1$ in the valley $K$ ($\xi = +1$) and $B2$, $A1$, $B1$, $A2$ in the valley $K'$ ($\xi = -1$). In the present study, we neglect the tight-binding parameters $\gamma_3$ (which leads to a trigonal warping of the band structure at low energies in Fig.1(b)) and $\gamma_4$ corresponding to the renormalization of $\gamma_1$.

The role of different kinds of gaps in the magneto-transport properties of bilayer may be studied for low-energy approximation, for energy range $|\epsilon| < \gamma_1/4$ may be studied [9] within 2-band model with (2x2) Hamiltonian:

$$H_{eff} = \xi \begin{pmatrix} \Delta_{\xi s} & -\xi(\hat{\pi}^+)^2/2m \\ -\xi \hat{\pi}^2/2m & -\Delta_{\xi s} \end{pmatrix} \quad \text{with effective mass} \quad m = \gamma_1/(2v_F^2) = 0{,}032 m_e \quad (3)$$

which is obtained from 4-band Hamiltonian (2) by integrating out $B1$, $A2$ fields which correspond to the Bernal stacked orbitals with neglecting $\gamma_3$, $\gamma_4$ hopping parameters. In turn, $H_{eff}$ acts on a two-component wave function $\Psi_2$ corresponding to the atomic sites $A1$, $B2$ in the valley $K$ ($\xi = +1$) and $B2$, $A1$ in the valley $K'$ ($\xi = -1$).

The Hamiltonians possess by the reflection properties in depending on the form of the gap. Thus for $B=0$ the combined effective Hamiltonian with the gap $\Delta_{\xi s} = U$ for two valleys, $H(\mathbf{p},U) = H_{eff}(\xi=+1, \mathbf{p}, U) \oplus H_{eff}(\xi=-1, \mathbf{p}, U)$ is time-reversal invariant under the transformation $(\Pi_1 \otimes \tau_1) H(\mathbf{p},U) (\Pi_1 \otimes \tau_1) = H(-\mathbf{p}, U)$, where $\Pi_1$ swaps $\xi = +1$ and $\xi = -1$ in valley space, while $\tau_1$ acts as time reversal. Whereas in the 4 × 4 case (2) the presence of the gap $\Delta_{\xi s} = \xi \Delta_T$ breaks the time-reversal symmetry. Indeed, for $B \neq 0$ we have: $(\Pi_1 \otimes \tau_1) H(\mathbf{p}, \Delta_T, B)(\Pi_1 \otimes \tau_1) = H(-\mathbf{p}, -\Delta_T, -B)$.

## OPTICAL CONDUCTIVITIES IN AND WITHOUT MAGNETIC FIELD

To calculate the optical conductivities in analytic way we use the Kubo formula with retarded current-current correlation function $\Pi_{kl}^R(\Omega + i0)$,

$$\sigma_{kl}(\Omega) = (\hbar/i\Omega)\left(\Pi_{kl}^R(\Omega+i0) - \Pi_{kl}^R(0)\right), \text{ for } \Pi_{kl}^R(i\Omega) = \Pi_{kl}(i\Omega_m \to \Omega + i\varepsilon), \; k,l = x,y \quad (4)$$

obtained by the analytical continuation from its imaginary time expression, and $\Omega$ is the energy of photon. The calculation of the current-current correlation function in tree level approximation reduces to the evaluation of the bubble diagram

$$\Pi_{kl}(i\Omega_m) = -\frac{1}{V}\int_0^\beta d\tau\, e^{i\Omega_m \tau}\iint d^2r\, d^2r'\, tr_{\xi,s}[j_k(r)G(r,r',\tau)j_l(r')G(r',r,-\tau)] \text{ with } j_k(r) = -c\,\partial H/\partial A^k \quad (5)$$

being by the electric current density operator. The quantities $G(r, r', \tau)$, $V$, $\beta = 1/T$, $\Omega_m = 2\pi m/\beta$ are the electron Green's function (GF), the volume of the system, the inverse temperature, Matsubara frequences and $tr_{\xi,s}$ implies a summation over 2x2 or 4x4 matrices and the valley and spin indices.

Note, in the presence of a magnetic field the GF being by inverse operator matrix for corresponding Hamiltonians (2), (3) is not translational invariant and calculations in both 2- and 4-band cases not so trivial due to Landau levels appearance in the energy spectrum for bilayer whereas for $B=0$, the GF's is translation invariant and due to dispersion relations among the energy and momenta we may to calculate (5) on a base of frequency-momentum representation

For the 2-band model in a magnetic field the optical conductivities $\sigma_\pm(\Omega) = \sigma_{xx}(\Omega) \pm i\sigma_{xy}(\Omega) = (\hbar/i\Omega)(\Pi^R_\pm(\Omega) - \Pi^R_\pm(0))$ corresponding to the opposite circular polarizations of light have the form (for magnetic length $l = \sqrt{hc/eB}$ [9]),

$$\sigma_\pm(\Omega) = e^2\hbar^2/2\pi m^2 l^4 \sum_{k\geq 0}(k+1)\sum_{\nu,\nu',\xi,s=\pm}\left[\frac{1}{\nu'M_{k+2} - \nu M_{k+1}}\left(1 - \frac{\nu\nu'\xi\Delta_{\xi s}}{M_{k+1}M_{k+2}}\right) \pm \frac{\nu\nu'\xi\Delta_{\xi s}}{M_{k+1}M_{k+2}}\right]\frac{i(n_F(\nu M_{k+1}) - n_F(\nu'M_{k+2}))}{\Omega \mp \nu'M_{k+2} \pm \nu M_{k+1} + 2i\Gamma} \quad (6)$$

for finite Landau levels width $\Gamma$, Fermi-Dirac distribution function, $n_F(\omega) = [\exp(\{\omega - \mu\}/T) + 1]^{-1}$ and energy Landau levels $E_{n\xi}$ for $\Delta_{\xi s} = \Delta_\xi = U + \xi\Delta_T$ determined from the spectral problem for the stationary Schrodinger equation $H_{eff}\Psi_{2n} = E_{n\xi}\Psi_{2n}$: $E_{n\xi} = \xi\Delta_\xi, n = 0,1$ $E_{\pm n\xi} = \pm M_{n\xi}, M_{n\xi} = \sqrt{\Delta_\xi^2 + \omega_c^2 n(n-1)}, n \geq 2$ for cyclotron energy $\omega_c = e\hbar B/mc$.

The limiting cases of a magnetic field absence for the optical Hall conductivity ($\sigma_{xy}(\Omega) = (\sigma_+(\Omega) - \sigma_-(\Omega))/(2i)$) and its observable real part for direct current obtained from (6) for continuum limit ($M_k \to \omega$) and $\Gamma=0$ [9]

$$\sigma_{xy}(\Omega, B=0) = -4e^2/\hbar \sum_{\xi,s=\pm}\xi\Delta_{\xi s}\int_{|\Delta_{\xi s}|}^{\infty}d\omega\frac{(n_F(\omega) - n_F(-\omega))}{4\omega^2 - (\Omega + i0)^2}, \quad \mathrm{Re}\,\sigma_{xy}(\Omega \to 0) = e^2/\hbar\sum_{\xi,s=\pm}\xi\begin{cases}\mathrm{sgn}(\Delta_{\xi s}), & |\mu| < |\Delta_{\xi s}| \\ \Delta_{\xi s}/|\mu|, & |\mu| \geq |\Delta_{\xi s}|\end{cases} \quad (7)$$

do not vanish for time-reversal breaking gaps $U_T, \Delta_T$ presence and $\mathrm{Re}\,\sigma_{xy}(\Omega \to 0)$ for neutral point, $\mu=0$, corresponds to the quantum anomalous Hall effect in bilayer.

## OPTICAL HALL CONDUCTIVITY AND THE FARADAY, KERR ANGLES

In turn, for the 4-band model with the Hamiltonian (2) in the magnetic field the optical conductivities may be determined in the same way as for 2-band model. We restrict ourselves by only presentation of Hall $\sigma_{xy}(\Omega)$ conductivity:

$$\sigma_{xy}(\Omega) = e^2 V_F^2 \hbar/\pi l^2 \sum_{k\geq 0}\sum_{\alpha,\beta,\alpha',\beta';\xi,=\pm}\left[\frac{(n_F(E_{k+1,\alpha',\beta'}(\xi)) - n_F(E_{k,\alpha,\beta}(\xi)))}{E_{k,\alpha,\beta}(\xi) - E_{k+1,\alpha',\beta'}(\xi)}\right]|C^*_{1,k}(\alpha,\beta,\xi)C_{4,k+1}(\alpha',\beta',\xi) + C^*_{3,k}(\alpha,\beta,\xi)C_{2,k+1}(\alpha',\beta',\xi)|^2 \quad (8)$$

$$\times\left(\frac{1}{\Omega + i\Gamma + E_{k,\alpha,\beta}(\xi) - E_{k+1,\alpha',\beta'}(\xi)} - \frac{1}{\Omega + i\Gamma - E_{k,\alpha,\beta}(\xi) + E_{k+1,\alpha',\beta'}(\xi)}\right)$$

where the constants $C_{i,k}(\alpha,\beta,\xi)$ and Landau levels now for each $k$ has 4-band, not-degenerate in spin s and valley $\xi$ and determined from the exact solution for the spectral problem $H\Psi_{4k} = E_{k,\alpha,\beta}(\xi)\Psi_{4k}(s, C_{1k}(\alpha,\beta,\xi), C_{2k}(\alpha,\beta,\xi), C_{3k}(\alpha,\beta,\xi), C_{4k}(\alpha,\beta,\xi))$ in the form (for details see, [10]):

$$E_0(\xi) = \xi\Delta_{\xi s}, \quad E_{1,\alpha}(\xi) = -\tfrac{2}{3}\sqrt{3\gamma_1^2 + 3a^2 + 4\Delta_{\xi s}^2}\cos\left(\tfrac{1}{3}\arccos\left[\frac{\xi\Delta_{\xi s}(8\Delta_{\xi s}^2 + 9(\gamma_1^2 - 2a^2))}{\sqrt{(3\gamma_1^2 + 3a^2 + 4\Delta_{\xi s}^2)^3}}\right] + a\frac{2\pi}{3}\right) + \tfrac{1}{3}\xi\Delta_{\xi s}, \alpha = 0,\pm 1,$$

$$E_{k,\alpha,\beta}(\xi) = 0.5(-1)^\beta\sqrt{2z_0(k)} + 0.5(-1)^\alpha\left(-2(z_0(k) - \gamma_1^2 - a^2(2k-1) + 2\Delta_{\xi s}^2) + (-1)^\beta 2\xi a^2\Delta_{\xi s}\sqrt{2/z_0(k)}\right)^{-\tfrac{1}{2}}, \alpha,\beta = 1,2, k \vartriangleright 2 \quad , \quad (9)$$

$$\begin{pmatrix}C_{1k}(\alpha,\beta,\xi)\\C_{2k}(\alpha,\beta,\xi)\\C_{3k}(\alpha,\beta,\xi)\\C_{4k}(\alpha,\beta,\xi)\end{pmatrix} = \frac{1}{\sqrt{2\pi l A_{k,\alpha,\beta}}}\begin{pmatrix}-i\xi a\sqrt{k}[(\xi\Delta_{\xi s} + E_{k,\alpha,\beta})^2 - a^2(k-1)]\\-i\xi a\sqrt{k-1}\gamma_1(\xi\Delta_{\xi s} - E_{k,\alpha,\beta})\\\gamma_1(\Delta_{2\xi}^2 - E_{k,\alpha,\beta}^2)\\[(\xi\Delta_{\xi s} + E_{k,\alpha,\beta})^2 - a^2(k-1)](\xi\Delta_{\xi s} - E_{k,\alpha,\beta})\end{pmatrix}, \quad A_{k,\alpha,\beta} = [(\xi\Delta_{\xi s} + E_{k,\alpha,\beta})^2 - a^2(k-1)]^2[(\xi\Delta_{\xi s} - E_{k,\alpha,\beta})^2 + a^2 k] + a^2(k-1)\gamma_1^2(\xi\Delta_{\xi s} - E_{k,\alpha,\beta})^2 + \gamma_1^2(\Delta_{2\xi}^2 - E_{k,\alpha,\beta}^2)^2,$$

$$C_{i0}(\alpha,\beta,\xi) = \delta_{i1}/(2\pi l), \quad (C_{11},C_{21},C_{31},C_{41}) = (\sqrt{2\pi l A_{1,\alpha}})^{-1}(-i\xi a(\xi\Delta_{\xi s} + E_{1,\alpha}), 0, \gamma_1(\xi\Delta_{\xi s} - E_{1,\alpha}), (\Delta_{2\xi}^2 - E_{1,\alpha}^2))$$

for $a = \hbar v_F\sqrt{2}/l$ and $z_0(k)$ being by one from the positive roots for cubic equation (according to Ferrari method of solution for the quartic equation on the energy $E_{k,\alpha,\beta}(\xi)$, for $k \geq 2$)

The spontaneously broken time-reversal symmetry in bilayer can be observable via optical polarization rotation when light is transmitted through the sample (Faraday effect) or reflected by it (Kerr effect) among them for: (1) free-standing bilayer with a refractive index n=1 (the first inset from the right in Fig.3a); (2) graphene on a thick substrate (corresponding to $SiO_2$, BN) with $n=1.5$; (3) for bilayer on a dielectric layer with $n=1.5$ and a thickness of

$d$ = 300nm on top of a thick layer with $n_s$=3.5 (which corresponds to the most standardly used SiO2/Si substrates). The Faraday and Kerr angles are determined by the rule:

$$\theta_F = 0.5(\arg(t_-) - \arg(t_+)), \; \theta_K = 0.5(\arg(r_-) - \arg(r_+))$$
$$(r_\pm, t_\pm) = (1 + n + Z_0\sigma_\pm)^{-1}(1 - n - Z_0\sigma_\pm, \; 2) \qquad (10)$$

(for the reflection and transmission coefficients $r_\pm, t_\pm$ at the 'vacuum-film-substrate' interface, for an impedance of vacuum, $Z_0 = 4\pi/c$) which for the limited case, $|Z_0\sigma_\pm| \ll n-1$, reduced to the relations $(\theta_F, \theta_K) \approx Z_0 \operatorname{Re}\sigma_{xy}/[n-1](-2/n+1, \; 1)$ described in terms of Hall conductivity.

The dependence of the Faraday and Kerr angles on the frequency Ω for various samples is shown on Fig. 3, where for different samples the Faraday angle for the gap $\Delta_T$ have the same behavior, whereas for Kerr angle the 3-rd sample with double substrate has another dependence after 0.4eV than the rest other samples

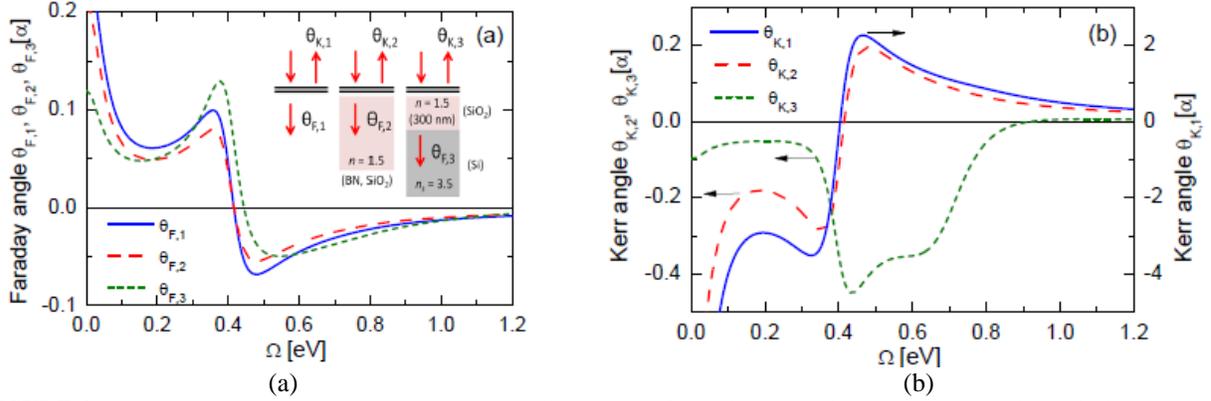

(a)      (b)

**FIGURE 3** The calculated Faraday (a) and Kerr (b) rotation angles for various experimental geometry shown in inset. $\sigma_\pm(\Omega)$ calculated within 2-band model for only non-vanishing time-reversal breaking gap $\Delta_T$ ($\Delta_T$=0,001eV) [9].

## SUMMARY

We have studied the influence of different kinds of gaps in bilayer graphene in the two- and four-band models on longitudinal and Hall optical conductivities in dependence of the choice of the gaps that break the time reversal symmetry. The two-band model is valid for energies $E$<100meV and the four-band model is applicable up to energies when continuum approximation is valid. The results may be applied for graphene-like optoelectronics.

## ACKNOWLEDGMENTS

The study of was partially supported by the grant of Leading Scientific Schools of the Russian Federation under Project No. 88.2014.2 and by the European IRSES Giant SIMTECH No.246937.